\documentclass[
    ,final            
  ]
  {aipproc}

\layoutstyle{6x9}

\begin{document}

\title{The evolutionary phase of B[e] supergiants and unclassified B[e] stars}

\author{Michaela Kraus}{
  address={Astronomical Institute, Utrecht University, Princetonplein 5, 3584 CC Utrecht, The Netherlands}
}



\begin{abstract}
We present two classes of stars with yet unknown evolutionary phase: the 
B[e] supergiants and the so-called unclassified B[e] stars. While the B[e]
supergiants are luminous post-main sequence stars with high mass progenitors,
not much is known of the unclassified stars. We discuss how it might be 
possible to determine their evolutionary phases in the area
of new large telescopes.
\end{abstract}

\maketitle


\section{Introduction}

Stars showing the B[e] phenomenon have several characteristics of their 
optical spectrum in common: they are of spectral type B, have strong Balmer 
emission lines, and many permitted and forbidden emission lines of 
predominantly low ionization metals like e.g. Fe{\sc ii} and O{\sc i}.
In addition, these stars are known to possess a strong near- or mid-infrared
excess due to hot circumstellar dust. 
Since most of the characteristics are based purely on the optical spectrum,
these stars can be of quite different evolutionary phase, and \citet{l1998}
grouped all stars with the B[e] phenomenon accordingly.
The different classes they found are: B[e] supergiants, Herbig Ae/B[e] stars, 
and compact planetary nebula B[e] stars. The biggest group, however, is 
formed by stars of unknown evolutionary phase, the so-called unclassified 
B[e] stars.
In this paper we want to concentrate on two classes of B[e] stars: the 
supergiants and the unclassified stars. Both classes have in common, that
their evolutionary phase is still unknown and we want to strengthen the 
need of large telescopes for the determination of the evolutionary phases of
these stars.

\section{The B[e] supergiants}

B[e] supergiants are luminous ($4 \leq \log L/L_{\odot} \leq 6$) post-main 
sequence objects. Their progenitors must have been massive ($7 < M/M_{\odot} 
< 85$) and probably rapidly rotating stars. The best known B[e] supergiants are
located in the Magellanic Clouds. In the Milky Way only some candidates have been 
found. Here, due to large interstellar extinction the proper determination of
distances and therefore luminosities is difficult.
The optical spectra of B[e] supergiants show a hybrid 
character, i.e. a co-existence of very broad Balmer lines and extremely narrow 
forbidden and permitted low ionization metal lines. This hybrid character has 
led to the interpretation of two different winds: a normal line-driven 
(CAK-type) fast wind of low density in polar direction and a slow, high density 
disk-forming wind in equatorial direction where the hot dust is located (see 
\citet{z1985}).
\citet{k2003} performed ionization structure calculations in such non-spherical
winds of B[e] supergiants using a latitude dependent mass flux that increases
from pole to equator. They found that with such a model hydrogen can recombine 
close to the stellar surface in the equatorial region leading to a hydrogen
neutral disk-like structure around these luminous stars. High-resolution 
optical spectra of B[e] supergiants revealed strong emission of O{\sc i} 
forbidden lines (see e.g. \citet{kb2004}) which indicates the existence of a 
huge amount of neutral hydrogen since H and O have the same ionization 
potential. The additional detection of CO band emission in the near-infrared in 
several B[e] supergiants (e.g. \citet{Mc1}; \cite{Mc2}; \cite{Mc3}) completes 
the picture of the circumstellar disk leading to the
following schematic representation of its temperature structure (from inside out): 
[O{\sc i}] emission (10\,000\,K $\leq T \leq$ 6\,000) -- CO band emission 
(5\,000\,K $\leq T \leq$ 2\,000) -- dust emission ($T \leq 2\,000$\,K).

In the HR diagram, B[e] supergiants share their location with other massive 
post-main sequence stars like Wolf-Rayet stars and Luminous Blue Variables 
(LBVs). But the connection of B[e] supergiants with these other evolutionary
states is still an unsolved problem and even newest stellar evolution
calculations of massive stars cannot predict the B[e] supergiant phase 
(see e.g. \citet{m2000}; \citet{m2004}).

\section{Unclassified B[e] stars: the example of Hen 2-90}

The enigmatic galactic star Hen 2-90 belongs to the group of unclassified
B[e] stars. It has been classified in the literature either as a compact
planetary nebula (e.g. \citet{c1993}; \citet{l1998}) or as a symbiotic object
(\citet{s2000}; \citet{g2001}). 
On an HST image (\citet{s2002}) a non-spherical wind structure is visible:
a high-ionized polar wind, a low-ionized wind at intermediate latitudes
and a (neutral) disk structure in equatorial direction.

We took high- and low-resolution optical spectra with slit centered
on this non-spherical wind structure. The spectra contain very strong Balmer
lines and lots of forbidden emission lines from e.g. S, N, O, Ar, Cl, Fe in
different ionization states. We detected neither emission from He{\sc 
ii} nor TiO absorption bands, which are both the main characteristics of 
a symbiotic object. In addition, we performed a detailed analysis of almost all 
observed forbidden emission lines and found that C and O are depleted, which
means that Hen 2-90 should be an evolved object. Our conlusions are therefore,
that Hen 2-90 is most probable a compact planetary nebula (\citet{k2004}).

There are, however, a few puzzling details that have to date not been solved:
(i) Hen 2-90 shows a jet like structure with many knots being ejected regularly
on both sides of the star and perpendicular to the disk. (ii) The velocity
profiles of the different emission lines hint to a much more complex 
structure of the circumstellar material than might be explained with a simple 
wind model. (iii) We also found 
from our modeling that N is depleted, which cannot be understood in terms
of a normal single star evolution so that probably a binary nature of this star
needs to be taken into account (\citet{kba2004}). More observations are certainly 
needed to disentangle the nature of this fascinating object.

\section{How can we determine evolutionary phases ?}

The determination of the evolutionary phases of B[e] supergiants and 
unclassified B[e] stars strongly depends on the interplay of theory 
with observations. Here, we want to mention some of the methods we are
using to fix stellar and circumstellar parameters necessary for the 
classification.

\begin{itemize}

\item {\bf Abundance determinations:}

The determination of elemental abundances is one key project when dealing
with stellar evolution. It is thereby necessary to determine 
abundances in the circumstellar matter as well as on the stellar surface.
The circumstellar material mirrors
the surface abundance at the time of matter ejection, while the surface
itself gives the actual abundance of the star. The surface abundance can be 
quite different from the circumstellar abundance if (i) the star was ejecting
its complete hydrogen rich outer layers leaving a He rich surface behind, or 
(ii) the star is rapidly rotating leading 
to rotationally induced mixing of the internal material which results in a
continuous surface enrichement of processed material. Since rapidly rotating
stars have high mass loss rates, the mixing leads to an abundance gradient
in the wind material.

\item {\bf Determination of the geometry and kinematics of the circumstellar matter:}

The geometry of the system is a crucial point, especially if no direct imaging
is available or possible. As discussed in the previous section, the appearance 
and strength of specific (e.g. forbidden) emission lines can hint to a 
non-spherical density distributions and even to neutral material close to the 
hot star. A detailed analysis of forbidden (and permitted) emission lines is 
needed to derive the ionization structure in non-spherical winds and disks.
Especially in the case of B[e] stars we know that in polar direction 
highly-ionized, line-driven winds of rather high velocity emanate, while from 
equatorial directions the narrow emission lines of low-ionized or even neutral 
metals are observed. Both types of profiles will be present in high-resolution 
spectra. 

Besides the atomic and ionic lines also molecular emission, like e.g. the CO 
bands can be used to determine the kinematics of the emitting gas. Especially
for several B[e] supergiants the CO bands which arise in the near-infrared 
have been observed. High-resolution observations of the CO $2\longrightarrow 0$
band head display the complete velocity information of the CO emitting gas.
This band head shows e.g. a red peak and a blue shoulder in case of rotation.
By modeling the high-resolution, high signal-to-noise $2\longrightarrow 0$ 
band head structure one can discriminate contributions coming from rotation 
and/or outflow (see e.g. \citet{k2000}). 

\item {\bf Determination of the mass loss history:}

The analysis of the forbidden emission lines can also be used to derive 
(non-spherical) mass fluxes and therefore the total mass loss rate of a star
as has been shown by \citet{k2004} for the unclassified B[e] star Hen 2-90.
Mass loss rates of the B[e] supergiants and unclassified B[e] stars can then 
be compared with those available in the literature for stars of different 
initial conditions and in different evolutionary phases to find agreements
and therefore a possible evolutionary stage of each star.

\end{itemize}

\section{Observations needed}

For the modeling of the CO bands high-resolution NIR spectra are needed.
PHOENIX, the new high-resolution near-infrared spectrograph at Gemini-South
is most suitable for these observations.

For the determination of terminal velocities, the mass loss history and 
the surface and circumstellar matter abundances we need mainly high-resolution
optical (and UV) observations. Since many of the unclassified B[e] stars are 
southern objects and most of the nowadays known B[e] supergiants 
are located in the Magellanic Clouds, the new South African Large Telescope 
(SALT) will be the ideal tool to guarantee the success of our projects.
Its major targets will be the Magellanic Clouds (see contributions by David 
and William in this volume) allowing us to retrieve the high-resolution 
optical spectra which are needed for a proper classification of B[e] stars.






\begin{theacknowledgments}
This research was supported by the Nederlandse Organisatie voor
Wetenschappelijk Onderzoek grant No.\,614.000.310.
\end{theacknowledgments}


\bibliographystyle{aipproc}   

\end{document}